\documentstyle[aps,twocolumn,epsfig]{revtex} 
 
\begin{document} 
\author{Ph. H\"{a}gler$^{1}$, R. Kirschner$^{2}$, A. Sch\"{a}fer$^{1}$, 
L. Szymanowski$^{1,3}$, O.V. Teryaev$^{4,5}$} 
\address{\medskip $^{1}$Institut f\"{u}r Theoretische Physik, Universit\"{a}t 
Regensburg,
D-93040 Regensburg, Germany\\ 
$^{2}$Institut f\"{u}r Theoretische Physik, Universit\"{a}t Leipzig,
D-04109 Leipzig, Germany\\ 
$^{3}$Soltan Institute for Nuclear Studies,
Hoza 69, 00689 Warsaw, Poland\\ 
$^{4}$CPht (UMR C7644), Ecole Polytechnique, Palaiseau, France\\ 
$^{5}$BLTP, JINR, 141980 Dubna, Russia} 
\title{Towards a solution of the charmonium production controversy:
$k_\perp$-factorization
versus color octet mechanism} 
\maketitle 
 
\begin{abstract}
The cross section of $\chi_{c J}$ hadroproduction is calculated in the $k_\perp$
-factorization approach. We find a significant contribution of 
the $\chi_{c1}$ state due to non-applicability of the Landau-Yang theorem 
because of off-shell gluons. 
The results are in agreement with data 
and in contrast to the collinear factorization
show a dominance of the color singlet part and a strong
suppression of the color octet contribution.
Our results could therefore lead to a
solution of the
longstanding controversy between the color singlet model and the 
color octet mechanism. 
\end{abstract} 
 
\date{20.09.2000} 
 
 
The production of heavy quarkonia received a lot of attention from both
theory and experiment in recent years. 
It is e.g. the most prominent signal in the search for the quark gluon 
plasma. Its usefullness is, however, questionable as long as the
charmonium production process is not understood.
For a 
review we refer to \cite{Sch94,BF,Bottom}. Originally heavy 
quarkonium production was described in the color singlet model (CSM) \cite 
{Baier,Guberina}. Calculations based on this model and standard 
collinear factorization show however disagreement with the experimental data. For 
example the next-to-leading order (NLO) QCD collinear results for direct $%
J/\Psi $ hadroproduction underestimate the measured cross section at 
Tevatron by a factor of $\approx$ 50 (see fig.4 in \cite{Abe95} and
Ref.\cite{Abe97}). 
The proposed solution to this strong discrepancy is the so called 
color-octet-mechanism (COM) \cite{BF95,BBL95}, according to which a 
color octet $q\overline{q}$-pair which has been produced at short distances 
can evolve into a physical quarkonium state by radiating soft gluons. The 
COM introduces uncalculable non-perturbative parameters, the color octet matrix 
elements, which 
have to be determined by a fit to the data \cite{CLI,CLII}. The inclusion of 
the COM into NLO QCD collinear calculations leads in the case of hadroproduction
to a reasonable agreement 
with experiment \cite{CLI,CLII}. In these calculations the color octet 
contribution dominates.
 
On the other hand up to now the COM suffers at least from two unsolved 
problems. When the, supposedly universal, color octet matrix elements
are applied to electroproduction of heavy quarkonium the 
theoretical predictions fail to describe the data \cite{C99}. Furthermore 
the results of the COM for polarized heavy quarkonium hadroproduction seem 
to be incompatible with recent data from Tevatron \cite{M99}. 
 
The longstanding discrepancy between the results based on 
the CSM together with collinear factorization and the experimental data 
shows up especially strongly in the $k_{\perp }$-dependent cross sections 
from Tevatron \cite{Abe95}. Thus one can wonder if the collinear 
approximation, in which in NLO the only transverse momentum of the produced 
quarkonium comes from an additional final state gluon, is suitable at all. 
 
The aim of our paper is to clarify this question by a study of $\chi _{cJ}$
hadroproduction within the $k_{\perp }$
-factorization approach, which takes the nonvanishing transverse 
momenta of the colliding t-channel gluons into account. 
Generically this corresponds to taking into account 
new regions of the phase space of the colliding gluons which
is mandatory for the description of hard processes in the Regge
region.

More precisely we 
calculate the production of $J/\Psi$'s originating from radiative $\chi _{cJ}$ 
decays. In a recent study \cite{HKSST00} of open $b\bar{b}$
hadroproduction we found that $k_{\perp }$
-factorization gives far better results than
 NLO collinear QCD calculations and we 
expect a similar improvement for heavy quarkonium production. The main 
ingredients of our calculations in \cite{HKSST00} are the unintegrated gluon 
distribution and the effective next-to-leading-logarithmic-approximation 
(NLLA) $q\overline{q}$ -BFKL production vertex which we use in this article 
as well. The projection of the heavy quark-antiquark pair onto the 
corresponding charmonium state is described in the standard way within the 
non-relativistic-quarkonium-model \cite{Guberina,Baier,CLI,CLII}. 
\begin{figure}[h] 
\centerline{\epsfig{file=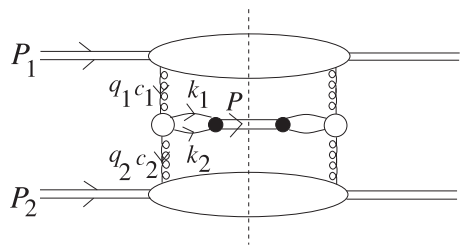,width=6cm}} 
\caption{The basic diagram} 
\label{cut} 
\end{figure} 
We study the production of $\chi _{cJ}$ whose lowest Fock 
state component is $q\overline{q}(^{3}P_{J})$. For $J/\Psi $ 
(a $q\overline{q}(^{3}S_{1})$ state) the LO production amplitude
is zero.
In order to get a nonzero $q
\overline{q}(^{3}S_{1})$-amplitude one has (in NLO in $\alpha _{S}$) to emit 
an additional gluon. The amplitude for the production of a $q\overline{q}$
-pair plus a gluon within the BFKL approach would in our case require an 
effective three-particle production vertex which still has to be
derived. 
In contrast the production of a $\chi _{c1}$ can be calculated in our
approach in LO 
because the Landau-Yang theorem which usually forbids the 
production of a $^{3}P_{1}$ state is not valid for off-mass-shell gluons.

We use the following definition of the light cone 
coordinates  
\[ 
k^{+}=k^{0}+k^{3},\text{ }k^{-}=k^{0}-k^{3},\text{ }k_{\perp 
}=(0,k^{1},k^{2},0)=(0,{\bf k,}0{\bf )}.  
\] 
In the c.m. frame the momenta of the scattering hadrons are given by  
\[ 
P_{1}^{+}=P_{2}^{-}=\sqrt{s},\text{ \ }P_{1}^{-}=P_{2}^{+}=P_{1\perp 
}=P_{2\perp }=0,  
\] 
where the Mandelstam variable $s$ is as usual the c.m.s. energy squared. The 
momenta of the t-channel gluons are $q_{1}$ and $q_{2}$ (see Fig.\ref{cut}).
The on-shell quark and antiquark (with mass $m$) have momentum $k_{1}$ 
respectively $k_{2}$ with  
\[ 
k_{1}^{-}=\frac{(m^{2}-k_{1\perp }^{2})}{k_{1}^{+}},\text{ \ }k_{2}^{-}=%
\frac{(m^{2}-k_{2\perp }^{2})}{k_{2}^{+}}.  
\] 
In the high energy (large $s$) regime we have  
\begin{eqnarray*} 
&&P^{+}\!=q_{1}^{+}-q_{2}^{+}\approx q_{1}^{+}, \;
P^{-}\!=q_{1}^{-}-q_{2}^{-}\approx -q_{2}^{-},\; 
q_{1/2}^{2}\approx q_{1/2\perp }^{2},
\end{eqnarray*} 
where $P=k_{1}+k_{2}$ is the momentum of the heavy quarkonium with $%
P^{2}=4m^{2}$. The longitudinal momentum fractions of the gluons are $%
x_{1}=q_{1}^{+}/P_{1}^{+}$, $x_{2}=-q_{2}^{-}/P_{2}^{-}$. 

The heavy quarkonium hadroproduction cross section in the $k_{\perp}$%
-factorization approach is \cite{CCH90}, \cite{CE91} 
 
\begin{eqnarray} 
&&\sigma _{P_{1}P_{2}\rightarrow \chi X} =\frac{1}{8(2\pi)}\!\!\int\!\!\frac{d^{3}P}{P^{+}}
d^{2}q_{1\perp }d^{2}q_{2\perp }  
\delta ^{2}(q_{1\perp }-q_{2\perp }-P_{\perp })
\nonumber\\ 
&&{\cal F}(x_{1},q_{1\perp }) 
\frac{1}{(q_{1\perp }^{2})^{2}} 
\left\{ \frac{\psi _{\chi }^{\dagger c_{2}c_{1}}\psi _{\chi }^{c_{2}c_{1}}
}{(N_{C}^{2}-1)^{2}}\right\} \frac{1}{(q_{2\perp }^{2})^{2}}{\cal F} 
(x_{2},q_{2\perp }).  \label{crosssection} 
\end{eqnarray} 
The factor $(N_{C}^{2}-1)^{2}$ comes from the projection on color 
singlet in the t-channel. ${\cal F}(x,q_{\perp })$ is the unintegrated gluon 
distribution. The heavy quarkonium production amplitude $\psi _{\chi 
}^{c_{2}c_{1}}(x_{1},x_{2},q_{1\perp },q_{2\perp },P)$ is 
factorized (see below) in a hard part which describes the production of the $%
q\overline{q}$ pair and an amplitude describing the binding of this pair 
into a physical charmonium state. We choose the scale $\mu ^{2}$ for 
$\alpha _{S}(\mu ^{2})$ in the amplitude $\psi 
_{\chi }^{c_{2}c_{1}}$ to be ${\bf q}_{1}^{2}=-q_{1\perp }^{2}$ 
respectively ${\bf q}_{2}^{2}=-q_{2\perp }^{2}$ \cite{LRSS91}. 
 
 
The amplitude for the production of the charmonium state can be written as  
\begin{eqnarray} 
&&\psi _{\chi }^{c_{2}c_{1}}={\cal P}(q\overline{q}\rightarrow \chi 
_{cJ})\bullet \Psi ^{c_{2}c_{1}}.  \label{chiamplitude} 
\end{eqnarray}
The $q\bar{q}$ production vertex $\Psi ^{c_{2}c_{1}}$ 
derived in \cite{FL96} for massless QCD, 
appropriately generalized for massive quarks, has the form  
\[ 
\Psi ^{c_{2}c_{1}}=-g^{2}\left( 
t^{c_{1}}t^{c_{2}}\,b(k_{1},k_{2})-t^{c_{2}}t^{c_{1}}\,b^{T}(k_{2},k_{1})%
\right) ,  
\] 
where $t^{c}$ are the colour group generators in the fundamental 
representation. The operator ${\cal P}(q\overline{q}\rightarrow \chi _{cJ})$ 
projects the $q\bar{q}$ pair onto the charmonium bound state, see below.
 
\begin{figure}[h] 
\centerline{\epsfig{file=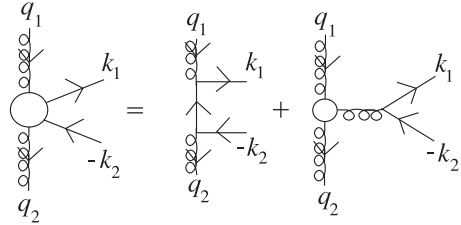,width=6cm}} 
\caption{The effective vertex} 
\label{flvertex} 
\end{figure}

The functions $b(k_{1},k_{2})$ and $b^{T}(k_{2},k_{1})$ are illustrated in 
Fig.\ref{flvertex} and their explicit form can be found in \cite{HKSST00}. 
One important property of the charmonium production amplitude for 
on-mass-shell quark and antiquark states (\ref{chiamplitude}), which is 
related to the gauge invariance of the whole approach, is its vanishing in the 
limit $q_{1\perp }\rightarrow 0$ (or $q_{2\perp }\rightarrow 0$). 
 
The relation between the usual gluon distribution $xg(x,{\bf q}^{2})$ and 
the unintegrated gluon distribution ${\cal F}(x,{\bf k})$ is given by 
 
\begin{eqnarray} 
&&xg(x,{\bf q}^{2})=\int_{0}^{\infty }\frac{d{\bf k}^{2}}{{\bf k}^{2}}\Theta 
({\bf q}^{2}-{\bf k}^{2}){\cal F}(x,{\bf k}).  \label{xg} 
\end{eqnarray} 
${\cal F}(x,{\bf k})$ includes the evolution in $x$ and ${\bf k}^{2}$ 
described by the BFKL and DGLAP equation. In the non-perturbative region of 
small ${\bf k}^{2}$ the unintegrated gluon distribution is not known, 
therefore we write (\ref{xg}) according to \cite{K85,CE91,RS94,RSS96} as  
\[ 
xg(x,{\bf q}^{2})=xg(x,{\bf q}_{0}^{2})+\int_{{\bf q}_{0}^{2}}^{\infty }%
\frac{d{\bf k}^{2}}{{\bf k}^{2}}\Theta ({\bf q}^{2}-{\bf k}^{2}){\cal F}(x,%
{\bf k}),  
\] 
which introduces the a priori unknown initial scale ${\bf q}_{0}$ and the 
initial gluon distribution $xg(x,{\bf q}_{0}^{2})$. Following \cite{K85,RS94}, 
we neglect the momentum dependence of the hard cross section 
in the soft region $|{\bf q}|<|{\bf q}_{0}|$, so that  
\begin{eqnarray*} 
&&\frac{1}{q_{1\perp }^{2}}\left\{ \frac{\psi _{\chi }^{\dagger 
c_{2}c_{1}}\psi _{\chi }^{c_{2}c_{1}}}{(N_C^{2}-1)^{2}}\right\} \frac{1}{ 
q_{2\perp }^{2}}\equiv S(q_{1\perp },q_{2\perp })\to \\ 
&&\left[S(q_{1\perp },q_{2\perp })\Theta (
{\bf q}_{2}^{2}-{\bf q}_{0}^{2}) 
+S(q_{1\perp },0)\Theta ({\bf q} 
_{0}^{2}-{\bf q}_{2}^{2})\right]\Theta ({\bf q}_{1}^{2}-{\bf q}_{0}^{2}) \\ 
&&+\left[S(0,q_{2\perp })\Theta ({\bf q}_{2}^{2}-{\bf q}_{0}^{2}) 
+S(0,0)\Theta ({\bf q}_{0}^{2}- 
{\bf q}_{2}^{2})\right]\Theta ({\bf q}_{0}^{2}-{\bf q}_{1}^{2}), 
\end{eqnarray*} 
see also the discussion of this expression in \cite{HKSST00}. 
 
One important point is the proper choice of the unintegrated gluon 
distribution function. We use the results of Kwiecinski, Martin and
Sta\`sto \cite{KMS97}. They determined it using a combination of DGLAP and BFKL 
evolution equations. With the initial 
conditions 
\begin{eqnarray} 
&&{\bf q}_{0}^2=1\text{ GeV},\ xg(x,{\bf q}_{0}^{2})=1.57(1-x)^{2.5}. 
\label{inivals} 
\end{eqnarray} 
they obtained an execellent fit to $F_{2}(x,Q^{2})$ data over a large 
range of $x$ and $Q^{2}$. 
 

In order to see the effect of off-shell gluons and the
inapplicability
of the Landau-Yang theorem as well as to perform calculations which do not
require a fit to the data we start with calculation of the color
singlet part of the amplitude.
This is most easily done by  
adapting the method of \cite{Baier,Guberina}. The projection 
of the hard amplitude onto the charmonium bound state is given by  
\begin{eqnarray} 
&&\psi _{\chi }^{c_{2}c_{1}} ={\cal P}(q\overline{q}\rightarrow \chi 
_{cJ})\bullet \Psi ^{c_{2}c_{1}}  \nonumber \\ 
&&=\sum_{i,j}\sum_{L_{z},S_{z}}\frac{1}{\sqrt{m}}\int \frac{d^{4}q}{(2\pi 
)^{4}}\delta \left( q^{0}-\frac{\vec{q}^{2}}{M}\right)\Phi
_{L=1,L_{z}}(\vec{q}) 
\nonumber \\ 
&&\left\langle 
L=1,L_{z},S=1,S_{z}|J,J_{z}\right\rangle
\left\langle 3i,\bar{3}j|1\right\rangle Tr\left\{ \Psi _{ij}^{c_{2}c_{1}}
{\cal P}_{S=1,S_{z}}\right\} ,  
\nonumber\\
\label{amplitude} 
\end{eqnarray} 
where $\Phi _{L=1,L_{z}}(\vec{q}=\vec{k}_{1}-\vec{k}_{2})$ is the momentum 
space wave function of the charmonium, and the projection operator ${\cal P}%
_{S=1,S_{z}}$ for a small relative momentum $q=k_{1}-k_{2}$ has the form 
 
\[ 
{\cal P}_{S=1,S_{z}}=\frac{1}{2m}(\not{k}_{2}-m)\frac{\not{\epsilon}(S_{z})}{%
\sqrt{2}}(\not{k}_{1}+m).  
\] 
The Clebsch-Gordan coefficient in color space is given by $\left\langle 3i,%
\bar{3}j|1\right\rangle =\delta _{ji}/\sqrt{N_{C}}$. Since $P$-waves vanish 
at the origin, one has to expand the trace in (\ref{amplitude}) in a Taylor 
series around $\vec{q}=0$. This yields an expression proportional to  
\[ 
\int \frac{d^{3}\vec{q}}{(2\pi )^{3}}q^{\alpha }\Phi _{L=1,L_{z}}(\vec{q})=-i%
\sqrt{\frac{3}{4\pi }}\epsilon ^{\alpha }(L_{z}){\cal R}^{\prime }(0),  
\] 
with the derivative of the $P$-wave radial wave function at the origin 
${\cal R}^{\prime }(0)$ whose 
numerical values can be found in \cite{EQ95}. For the 
individual $\chi _{cJ=1}$ and $\chi _{cJ=2}$ amplitudes we use 
\begin{eqnarray*} 
&&\sum_{L_{z},S_{z}}\!\!\left\langle 1,L_{z},1,S_{z}|1,J_{z}\right\rangle 
\epsilon ^{\mu }(L_{z})\epsilon ^{\nu }(S_{z})
\!=\!-i\sqrt{\frac{1}{2}}\varepsilon ^{\mu \nu \alpha \beta }\frac{P_{\alpha }
}{M}\epsilon _{\beta }(J_{z}) \\ 
&&\sum_{L_{z},S_{z}}\!\!\left\langle 1,L_{z},1,S_{z}|2,J_{z}\right\rangle 
\epsilon ^{\mu }(L_{z})\epsilon ^{\nu }(S_{z}) 
=\epsilon ^{\mu \nu }(J_{z}) 
\end{eqnarray*} 
where we introduce the spin 1 and spin 2 polarization tensors $\epsilon 
_{\beta }(J_{z})$ and $\epsilon ^{\mu \nu }(J_{z})$ of the produced 
charmonium $\chi _{cJ=1}$ respectively $\chi _{cJ=2}$. In the unpolarized 
case the squared amplitudes are further evaluated using  
\begin{eqnarray*} 
\sum_{J_{z}}\epsilon ^{\mu }(J_{z})\epsilon ^{\nu }(J_{z}) &=&-g^{\mu \nu }+ 
\frac{P^{\mu }P^{\nu }}{M^{2}}=P^{\mu \nu }, \\ 
\sum_{J_{z}}\epsilon ^{\mu \nu }(J_{z})\epsilon ^{\alpha \beta }(J_{z}) &=& 
\frac{1}{2}\left( P^{\mu \alpha }P^{\nu \beta }+P^{\nu \alpha }P^{\mu \beta 
}\right) -\frac{1}{3}P^{\mu \nu }P^{\alpha \beta }. 
\end{eqnarray*} 
The cross section for $J/\Psi $ production from radiative $\chi _{cJ}$ 
decays is then given by \cite{CLI,CLII}  
\[ 
\sigma _{J/\Psi \ from\ \chi _{c}}=\sum_{J=0,1,2}\sigma 
_{P_{1}P_{2}\rightarrow \chi _{cJ}X}\cdot Br(\chi _{cJ}\rightarrow J/\Psi 
+\gamma ),  
\] 
with the $\chi _{cJ}$ hadroproduction cross section $\sigma 
_{P_{1}P_{2}\rightarrow \chi _{cJ}X}$ (\ref{crosssection}). Because of the 
small branching ratio $Br(\chi _{cJ=0}\rightarrow J/\Psi +\gamma )={\cal O}%
(10^{-3})$ the contribution from $\chi _{cJ=0}$ is negligible. 
For the numerical computation we use the values  
\begin{eqnarray*} 
&&m_{c} =1.48\text{ GeV,}\;\;
|{\cal R}^{\prime }(0)|^{2} =0.075\text{ GeV}^{5}.
\end{eqnarray*} 
The pseudorapidity is defined as $\eta =\frac{1}{2}\ln \left( (\sqrt{ 
P_{0}^{2}-M^{2}}+P_{3})/ (\sqrt{P_{0}^{2}-M^{2}}-P_{3})\right) $.
To compare with data 
 we multiply our cross sections with the braching ratio  
$Br(J/\Psi \rightarrow \mu ^{+}\mu ^{-})$.
\begin{figure}[h]
\centerline{\epsfig{file=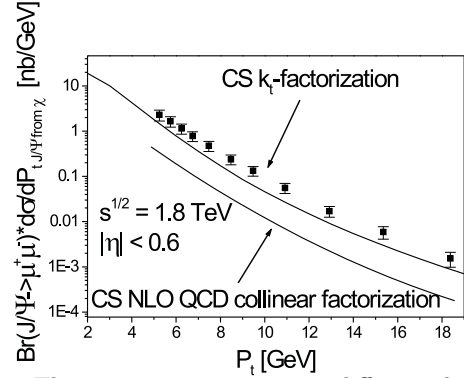,width=6cm}}
\caption{The transverse momentum differential cross section in 
comparison to the data and a 
NLO QCD calculation}
\label{chi}
\end{figure}
The resulting $P_{\perp }$%
-dependent cross section for $J/\Psi$'s from radiative deacays of $\chi _{c}$'s 
produced in $pp$-collisions is shown in Fig.\ref{chi} together with the 
 data from the CDF Collaboration \cite{Abe95} and a NLO 
QCD collinear result (see Fig.7 in \cite{CLII}). The individual 
contributions from $\chi _{c1}$ and $\chi _{c2}$ are shown in Fig.\ref 
{chi1chi2}.
\begin{figure}[h] 
\centerline{\epsfig{file=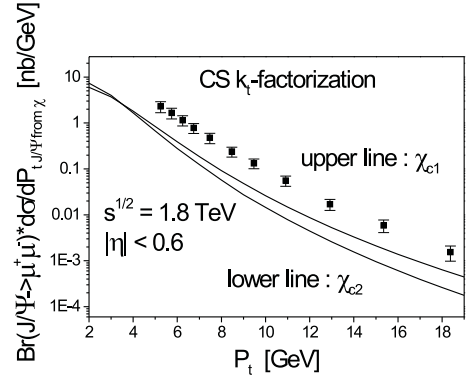,width=6cm}} 
\caption{The individual contributions from $\chi_{c1}$ and $\chi_{c2}$} 
\label{chi1chi2} 
\end{figure} 
The description of the data by the color singlet part alone
is very satisfactory and becomes even better 
if the difference of the transverse momentum of $J/\Psi $ (which is measured 
experimentally) and $\chi _{c}$ (which enters our calculation) is taken 
into account. (Due to the radiative decay the transverse momentum of $J/\Psi $ 
is typically larger by an amount of $\approx 300$ MeV than the corresponding  
$\chi _{c}$ one which leads to a shift of the theoretical curve to the right.) 
The typical scale of the gluon off-shellness is given by the
transverse momentum of the produced quarkonium. 

We emphasize that the result has been obtained without fitting any of the 
parameters involved: The unintegrated gluon distribution has been adopted 
from Kwiecinski et al. \cite{KMS97}. The parameters of the 
quarkonium bound state are 
the ones given by Eichten and Quigg \cite{EQ95}. 
For the $\chi _{c1}$ state it is
crucial that the gluons are off-shell in $k_{\perp }$ 
-factorization. 
 \begin{figure}[h]
\centerline{\epsfig{file=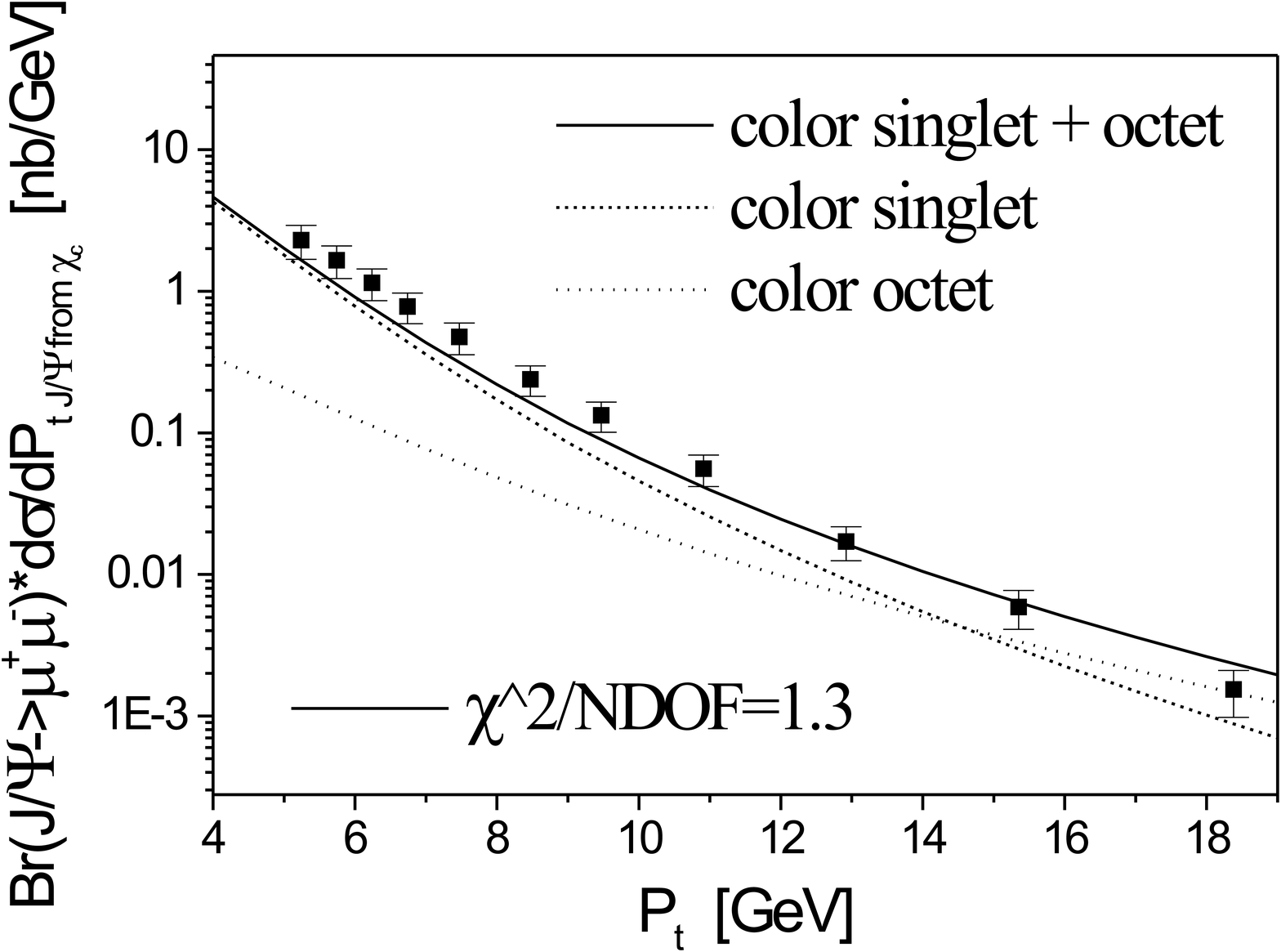,width=6cm}}
\caption{The color octet contribution}
\label{chioctet}
\end{figure}
Now we proceed with the calculation 
of $J/\Psi $ production by $\chi _{c}$ 
radiative decays adopting the colour octet mechanism. 
The infrared stability of higher order corrections to the cross
section
requires the existence of a color octet
contribution, without fixing its size
\cite{BBL95}.
The $\chi _{c}$ 
state can be written in a velocity expansion as \cite{CLI}  
\[ 
\left| \chi _{c}\right\rangle ={\cal O}(1)\left| q\overline{q}\left[ 
^{3}P_{J}^{\text{ }1}\right] \right\rangle +{\cal O}({\bf v})\left| q%
\overline{q}\left[ ^{3}S_{1}^{8}\right] g\right\rangle +\cdot \cdot \cdot . 
\] 
Following the formalism of \cite{CLI,CLII} the resulting cross section is 
then proportional to the color octet matrix element  
$\left\langle 0\right| O_{8}^{\chi _{c1}}(^{3}S_{1})\left| 0\right\rangle$ 
which has to be fitted to data. 
Using the results for the color singlet part and adding the color
octet
contribution we obtain as value of the color octet matrix element
$
\left\langle 0\right| O_{8}^{\chi _{c1}}(^{3}S_{1})\left|
0\right\rangle
=(9.0 \pm 2.0) \times 10^{-4}.
$ 
Comparing this with the result obtained in the 
collinear factorization \cite{CLI,CLII} we
find a suppression of the matrix element 
due to the flat $P_\perp$-dependence of the color octet
contribution by
roughly one order of magnitude, resulting in a
violation of the velocity scaling rules.
These scaling rules are derived rigorously in the framework
of non-relativistic QCD (NRQCD) \cite{BBL95}. 
It is, therefore, natural to assume that the charm
quark is simply not heavy enough for the velocity scaling rules
of NRQCD to be valid. This is also
suggested by other observations, see e.g. the very recent study
\cite{FRL}.
In contrast the description of bottom systems in NRQCD
should be more accurate.
This shows the importance of a detailed 
analysis of bottomonia production
in the $k_\perp$-factorization approach.


Let us conclude. The $k_{\perp }$-factorization approach relying on an 
unintegrated gluon distribution compatible with the small $x$ behaviour of 
the structure function $F_{2}$ together with the BFKL NLLA fermion 
production vertices describes correctly $\chi _{c}$ production in the 
central rapidity region. 
Whereas the standard collinear factorization 
approach in NLO can describe the data in the TeV range only by introducing 
a dominant octet contribution, we have shown that in the $k_{\perp }$-factorization approach 
 such a contribution gives an improved description of the
data but is suppressed by its $
P_{\perp }$ behaviour.  
 
 
Our main conclusion is therefore that the correct way to improve the 
standard QCD calculations for quarkonium production
in the TeV range is to abandon the 
collinear approximation. The contributions 
disregarded in the collinear approximation of strong transverse momentum 
ordering become essential in the small-$x$ range.
The relative merits of the $k_{\perp }$
-factorization as the standard approach for other processes in high
energy hadronic collisions still has to be investigated. 


 
L.Sz. and O.V.T. thank A. Tkabladze for discussion.
 
The work is supported by the DFG.

\end{document}